\documentclass[fleqn,11pt]{SelfArx} 

\usepackage[english]{babel} 

\definecolor{color1}{RGB}{0,0,90} 
\definecolor{color2}{RGB}{0,20,20} 

\usepackage{hyperref} 

\hypersetup{
	hidelinks,
	colorlinks,
	breaklinks=true,
	urlcolor=color2,
	citecolor=color1,
	linkcolor=color1,
	bookmarksopen=false,
	pdftitle={Title},
	pdfauthor={Author},
}

\usepackage[style=authoryear,
    backend=biber,
    maxbibnames=99,     
    dashed=false        
]{biblatex}
\JournalInfo{Accepted for publication in Solar Physics, 2025}
\Archive{DOI 10.1007/*************************}

\PaperTitle{North-South Asymmetry of the Solar Activity at Different Spatial Scales}
\PaperSubTitle{North-South Asymmetry of Solar Activity at Different Scales}

\Authors{V.N.~\ Obridko\textsuperscript{1}, 
A.S.~\ Shibalova\textsuperscript{1}, 
D.D.~\ Sokoloff\textsuperscript{1,2}*, 
I.M.~\ Livshits\textsuperscript{3}
}

\affiliation{\textsuperscript{1}\textit{IZMIRAN, 4, Kaluzhskoe Shosse, Troitsk, Moscow, 108840, Russia}}
\affiliation{\textsuperscript{2}\textit{Department of Physics, Lomonosov Moscow State University, Moscow, 119991, Russia}}
\affiliation{\textsuperscript{3}\textit{Sternberg Astronomical Institute, Lomonosov Moscow State University, Moscow, 119234, Russia}}
\affiliation{*\textbf{Corresponding author}: sokoloff.dd@gmail.com}

\newcommand{\keywordname}{Keywords} 

\Abstract{
Solar activity seems quite understandable when considered on the scales comparable 
with a solar cycle, i.e. about 11 years,  and on a short time scale of about a year. 
A solar cycle looks basically (anti)symmetric with respect to the solar equator, 
while the sunspot distribution is more or less random. We investigated the 
difference in the spatial distribution of magnetic structures on both time 
scales in terms of sunspots and the surface large-scale magnetic field and 
arrived at the conclusion that the structures of each type are created by a 
specific mechanism. For long-term structures, it is the mean-field dynamo. 
For the short-term ones, it is the spot production considered as a separate 
physical mechanism. The relationship between the mean-field dynamo mechanism 
and the processes of sunspot formation is a complex problem of current interest. 
The 11-year cycle itself is created by the mean-field dynamo and is most 
likely determined by processes in the convection zone. However, the 
transformation of magnetic flux into spots and active regions occurs, 
apparently, on significantly shorter time scales and probably develops 
directly in the subsurface layers, i.e., Near-Surface Shear Layer (NSSL) 
or leptocline.
}

\begin{document}

\maketitle 


\thispagestyle{empty} 


\section*{Introduction} 


When considering the solar activity cycle in the most general terms  \parencite[e.g.][]{U17}, people usually are speaking of two waves of activity in the form of sunspot groups of opposite polarity propagating from middle latitudes to the solar equator. The propagation takes about 11 years. Then, according to the Hale polarity law, new waves of polarity opposite to the one at the previous stage arise. The general structure is symmetric with respect to the solar equator. This idealized scheme is reproduced within the solar dynamo theory \parencite[e.g.][]{CS23}. A dipolar polar magnetic field is transformed by differential rotation into the antisymmetric toroidal field with respect to the solar equator. The poloidal magnetic field is restored from the toroidal field by some mirror-asymmetric factor, such as the meridional circulation and/or Leighton mechanism.

This symmetric scheme illustrates a situation that arises after the averaging is taken over a substantial timescale. Some north-south asymmetry survives even after averaging. 

In contrast, it seems quite difficult to see the north-south symmetry based on just a snapshot of the distribution of the solar activity tracers over the solar surface. Of course, there are days when the distribution mimics the symmetry somehow; however, it is more or less an exception. In other words, the properties of the solar-activity-cycle are related to average values and the mean-field dynamo problem, while snapshots and the local structure of the magnetic field are determined by some other physics.

The goal of our paper is to find the threshold separating these two regions, i.e. large-scale and small-scale behavior, and to see how one region transforms into the other as the scale changes.

To make our study more specific, we should note that the above-mentioned oversimplified scheme presumes a number of theoretical oversimplifications. The excitation mechanism common to both hemispheres is quite often understood as the identity of the dynamos acting in both hemispheres. The polar magnetic fields at both poles are expected to be identical. The polar magnetic field should be directly related to sunspot data and should be considered as the most suitable predictor of solar activity. All these expectations are to some extent exaggerations and should be adapted to the real observational situation. The latter obviously requires a critical approach since the dynamo considers the evolution of the large-scale magnetic field, whereas the relationship between the large-scale magnetic field and sunspots needs clarification. It seems plausible that the generation of the mean magnetic field is associated with general modulation of the solar activity tracers, but the scale at which the mechanism of formation of particular magnetic structures becomes important should still be identified.

In particular, the redistribution of large-scale magnetic flux by the large-scale negative effective magnetic pressure instability (NEMPI) leads to the formation of sunspots and active regions \parencite{Ketal89, KR90}. This instability has been studied theoretically using various analytical approaches \parencite{Ketal94,Ketal96, RK07} and was detected in direct numerical simulations \parencite[e.g.][]{Betal11, Betal16, Wetal16}. NEMPI has a threshold at a certain magnitude of the mean magnetic field. The authors claim that no new large-scale magnetic flux is produced in the process triggered by this instability. This interpretation is clearly at odds with the expectations arising from the usual interpretation of flux-transfer dynamos.

Note one more consideration that needs to be incorporated into the scheme. \textcite{Oetal02, Oetal03} show that the radial component of the magnetic flux at the outer boundary of the sunspot penumbra is about 550 $\mu$ s cm$^{-1}$, regardless of the spot area and the maximum magnetic field in the umbra. However, the relationship between the total flux and the spot area is substantially non-linear. An explicit parameterization of this relationship has been proposed to conclude that the contribution of sunspot-related magnetic flux to the total magnetic flux is small, not reaching more than 20\% even at solar maximum \parencite{Oetal02, Oetal03}.

\section{North-South Asymmetry Studies}

The North-South asymmetry of solar activity, discovered quite a long time ago, became the subject of more or less quantitative studies based on observations and/or dynamo context only in the last 10--20 years. Among the possible mechanisms responsible for the asymmetry, researchers have proposed various options, such as stochastic convective flows \parencite{Hetal94}, nonlinear back-reaction in dynamo models of the dynamo-driven magnetic field to hydrodynamic flows \parencite{W10}, or simply a relic magnetic field existing somewhere in the radiative core since the formation of the Sun \parencite{BL84, MK04}. However, the latter option seems dubious because the asymmetry varies substantially in sign and amplitude on both short and long timescales (see below). Another option is that the solar dynamo driven by differential rotation and mirror asymmetric flows acts quite independently in both hemispheres. While the dynamo in one hemisphere may become subcritical and the magnetic field may decay, in the other hemisphere it may remain supercritical and still produce there magnetic field  \parencite[e.g.][]{BD13}. The asymmetry in dynamo drivers may arise, in particular, due to the magnetic field back-reaction on the flow \parencite[e.g.][]{SN94, T97}. This anticorrelation is supported to some extent by \parencite{HW90, KN90, OS00a, OS00b, OS16} and somehow constrains the dynamo modeling. For a comprehensive review of the effects of the hemispheric interaction from the observational and theoretical points of view, see \parencite{VB90, Cetal93, Lietal02, Metal02, Betal05, Tetal06, Cetal07, Betal08, SR10,  BO11, Netal14}.

\section{Revising the North-South Asymmetry}

\begin{figure}    
\includegraphics[width=0.32\textwidth]{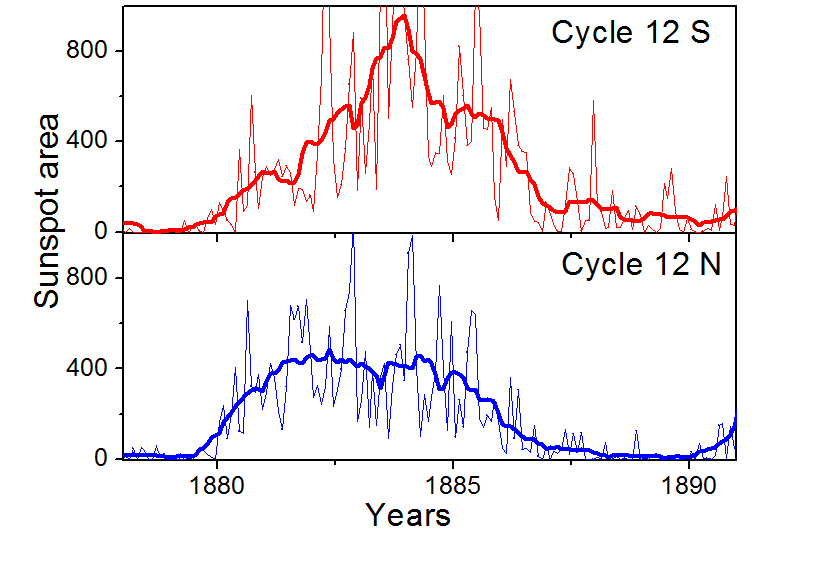}
\includegraphics[width=0.32\textwidth]{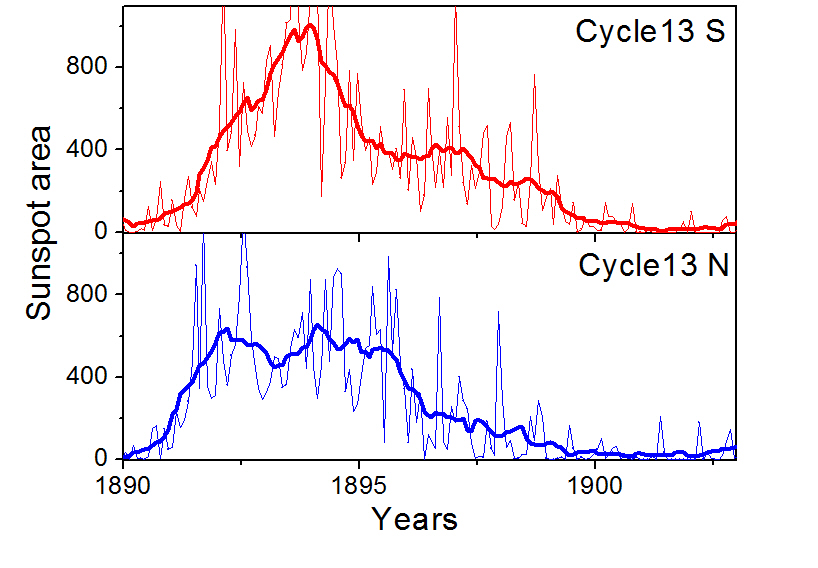}
\includegraphics[width=0.32\textwidth]{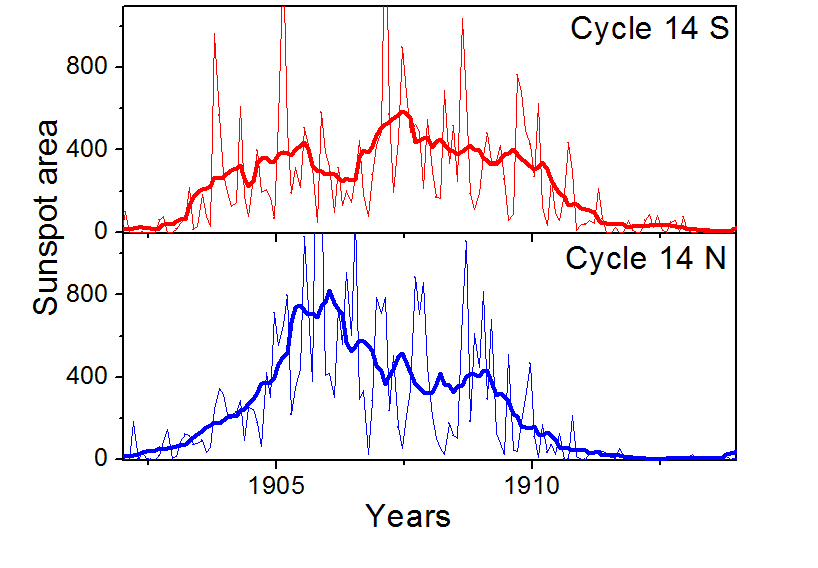}
\includegraphics[width=0.32\textwidth]{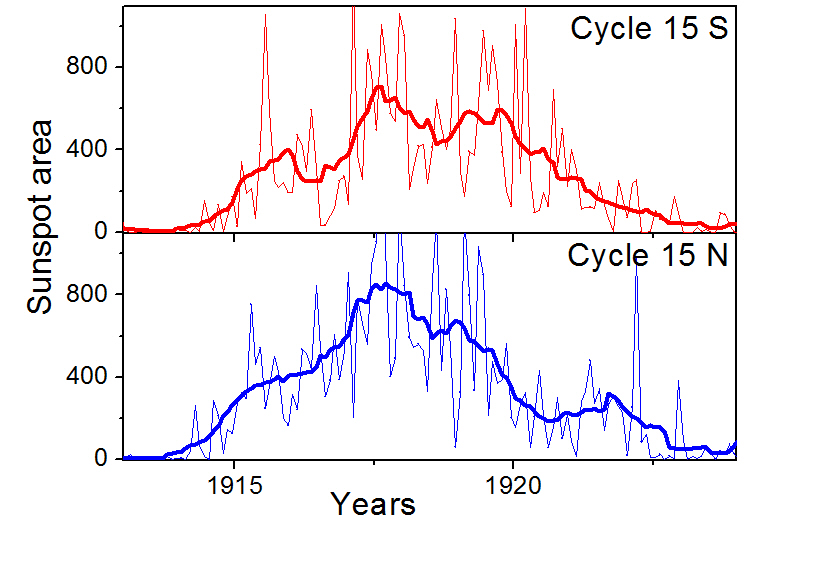}
\includegraphics[width=0.32\textwidth]{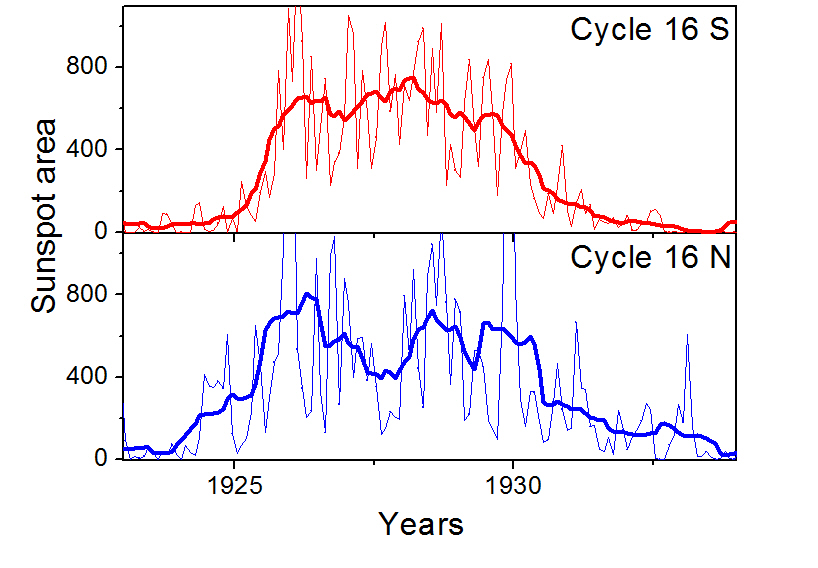}
\includegraphics[width=0.32\textwidth]{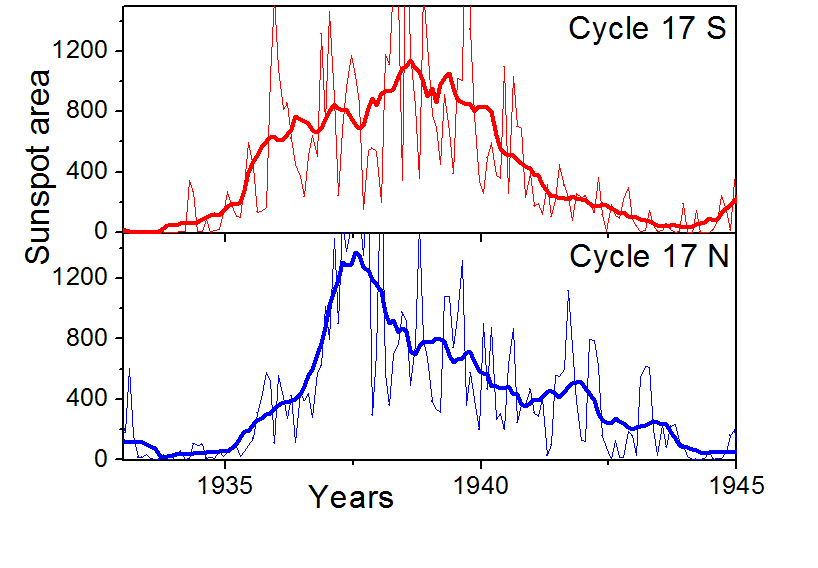}
\includegraphics[width=0.32\textwidth]{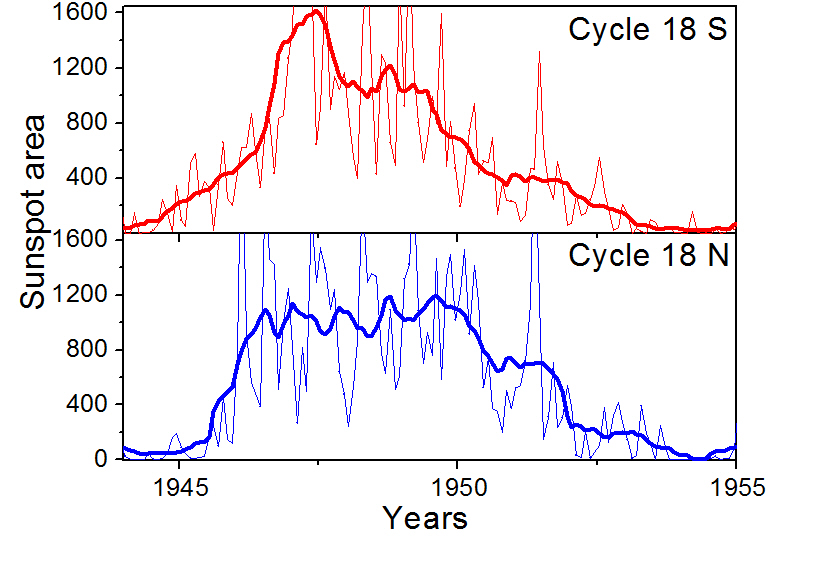}
\includegraphics[width=0.32\textwidth]{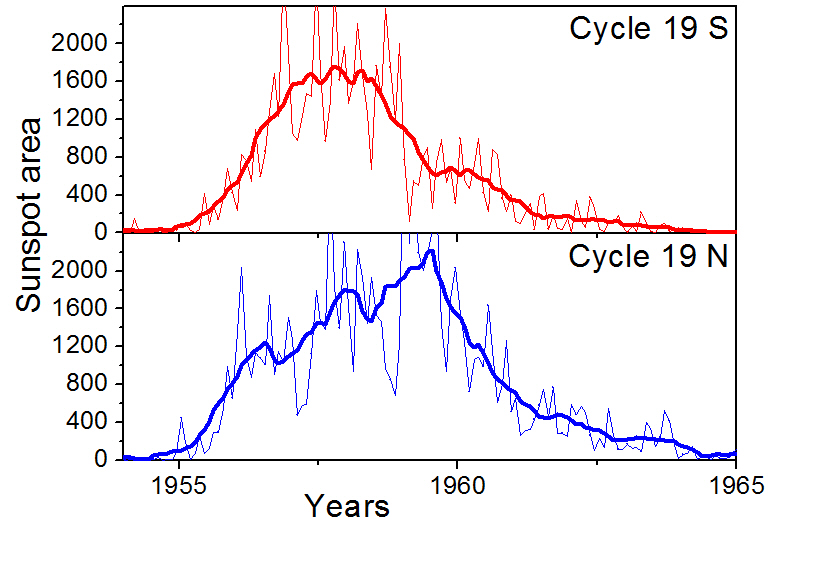}
\includegraphics[width=0.32\textwidth]{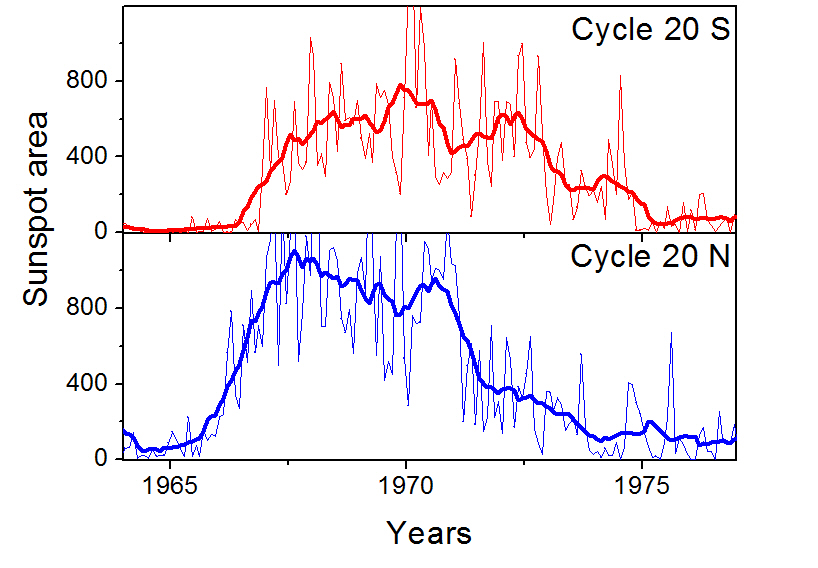}
\includegraphics[width=0.32\textwidth]{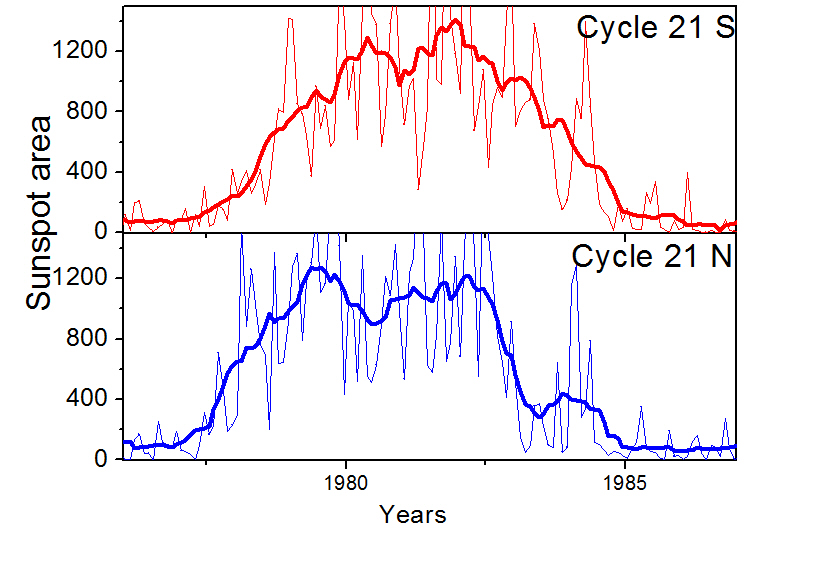}
\,
\includegraphics[width=0.32\textwidth]{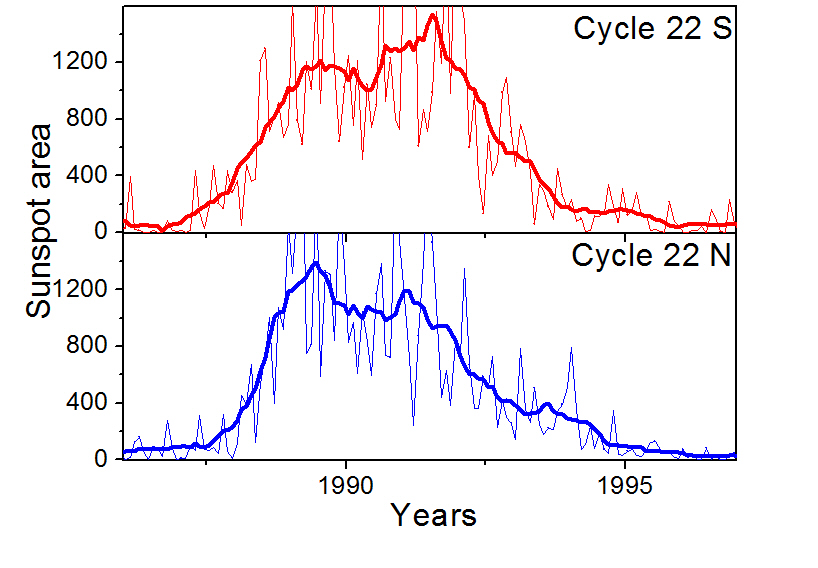}
\,
\includegraphics[width=0.32\textwidth]{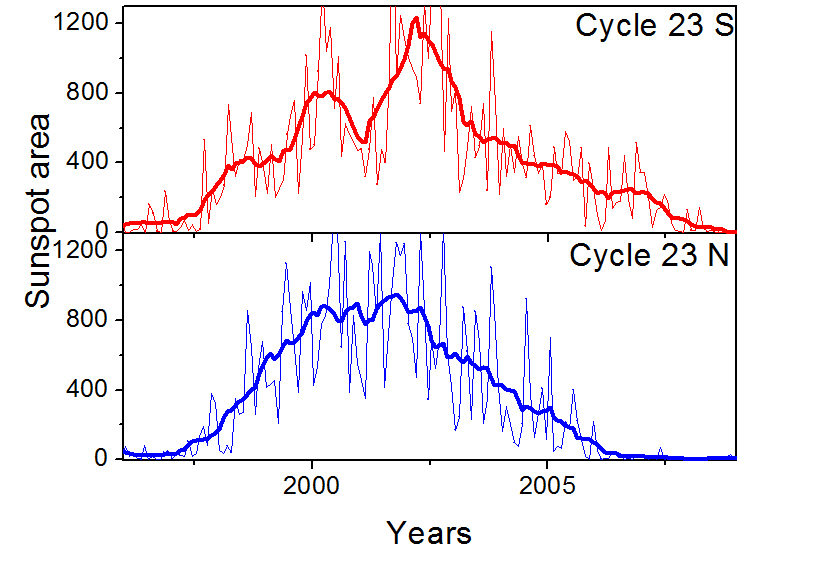}
\\
\small
        \caption{Cyclic variation of the sunspot areas (m.p.h.) in Cycles 12-23. The {\it red panels} show the 
 data for the south hemisphere; the {\it blue ones} stand for the north hemisphere. The {\it thin curves} demonstrate   monthly variations and the {\it thick ones}, the one-year smoothed data.
       }
\label{F1}
\end{figure}

Here we depart from the analysis performed in \textcite{Setal16} and revise data on the asymmetry of sunspot areas in Cycles 12--23 in the Greenwich database \url{solarscience.msfc.nasa.gov/greenwch.shtml} (Figure~\ref{F1}). Both hemispheres demonstrate an 11-year periodicity; however, the 11-year variations in the North and South hemispheres differ substantially. They start and finish not simultaneously and have a specific shape in the north and south hemispheres, which varies from cycle to cycle. It is difficult to claim that these are just synchronic harmonic oscillations. We will rather avoid hasty conclusions and will examine the asymmetry systematically from different points of view.

\begin{figure}    
\includegraphics[width=0.7\textwidth]{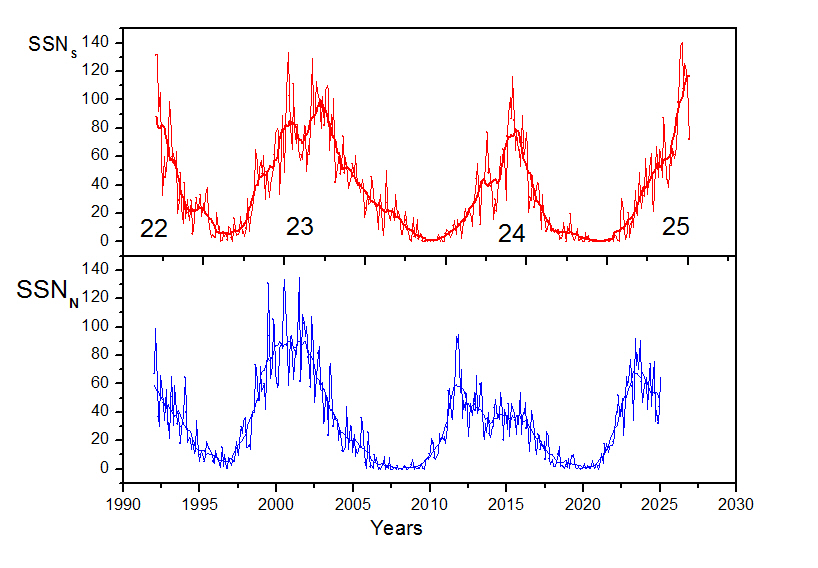}
\small
        \caption{Cyclic variation of sunspot numbers (SSN) in Cycles 22-25. {\it Red} stands for the south hemisphere and {it blue} stands for the north one. The {\it thin lines} illustrate the monthly data and the {\it thick ones},  the one-year smoothed data.
       }
\label{F2}
\end{figure}

Let us supplement Figure~\ref{F1}, based on the sunspot area data for Cycles 12--23, with Figure~\ref{F2}, based on the sunspot numbers for the past four cycles (22--25) in the new V2 system (\url{sidc.be/SILSO/datafiles}) (to save space, only Cycles 22-25 are shown).

\begin{figure}    
\includegraphics[width=0.32\textwidth]{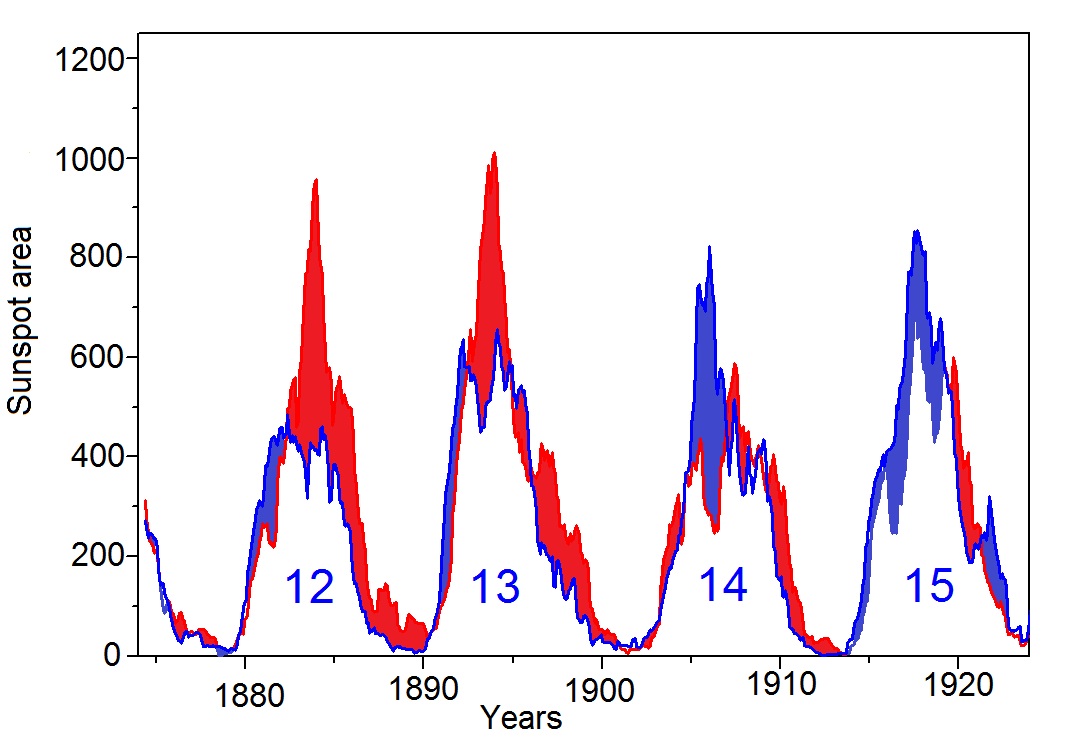}
\includegraphics[width=0.32\textwidth]{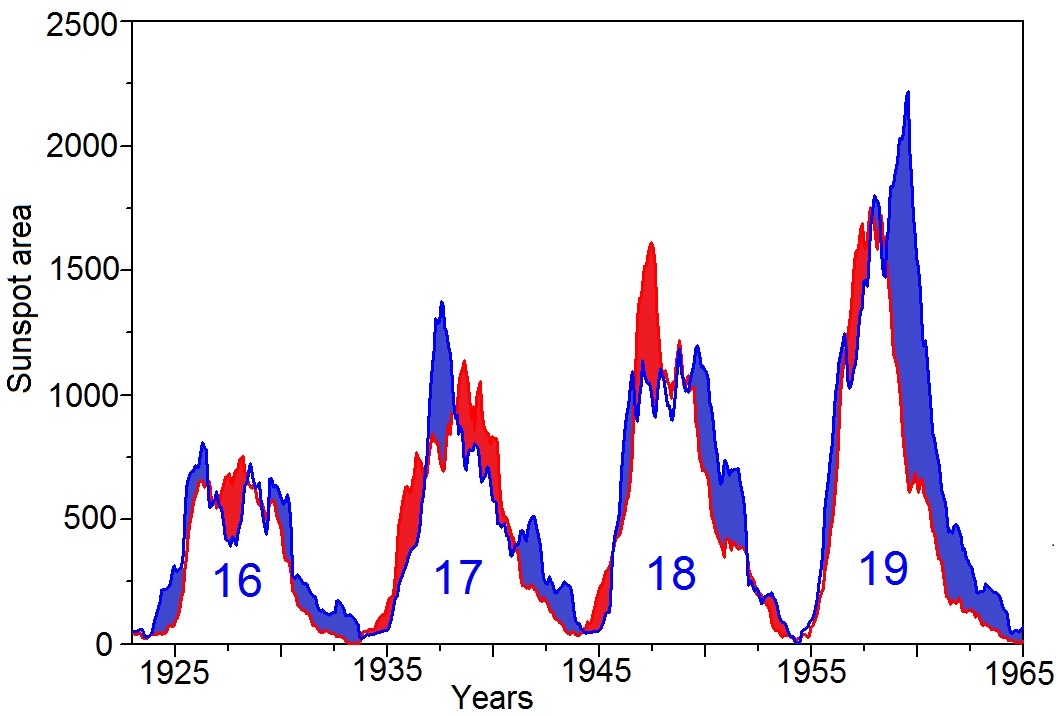}
\includegraphics[width=0.32\textwidth]{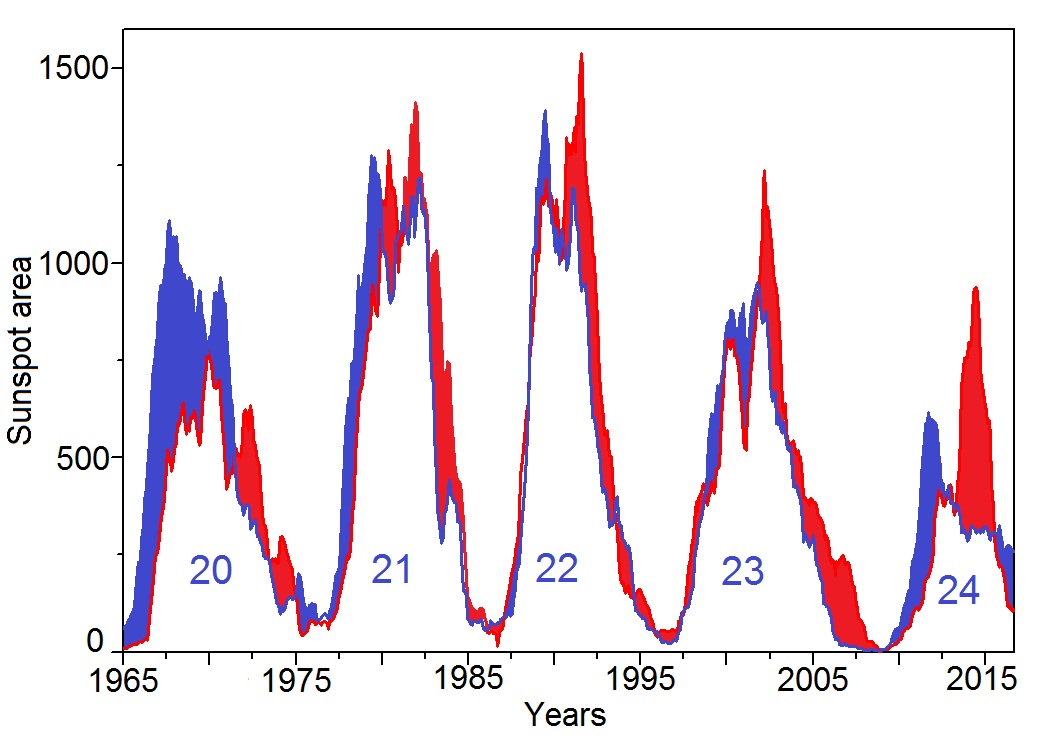}
\small
        \caption{Cyclic variation of the sunspot areas (m.p.h.) in Cycles 12-23. The data for the north and south hemispheres are overlaid. {\it Red} stands for the south hemisphere; {\it blue} stands for the north one. 
       }
\label{F3}
\end{figure}

It seems useful to superimpose the data for the northern and southern hemispheres (Figure~\ref{F3}).

Figures~\ref{F1}-\ref{F3} show that the development of cycles is not synchronized in different hemispheres. In Cycles 12-13 and (less pronounced) in Cycles 14-15, the northern hemisphere was in the lead. In Cycles 16-19, the southern hemisphere was in the lead, and in the last cycles the northern hemisphere was leading again. To summarize, we see that for about 60 years, the cycle starts with blue and then red becomes first again (Table 1). However, the variety of options is quite rich. Table 1 is an updated version of the corresponding table from \textcite{Setal16}. Data for the last two cycles are included based on \url{sidc.be/SILSO/datafiles}.

\begin{table}
\caption{Cycle maxima according to the sunspot area.}
    \begin{tabular}{|c|c|c|c|}
    \hline 
Cycle number & \multicolumn{2}{c|}{ Year of maximum} & Shift (months) \cr
 \cline{2-3}
         & North (yrs) & South (yrs) & \cr 
       \hline
         12 & 1882.5& 1883.3 & -10  \cr
         13 & 1893.3 & 1893.3 & 0 \cr
         14 & 1907.1 & 1907.3 & -3 \cr
         15 & 1917.0 & 1917.3 & -4 \cr
         16 & 1927.5 & 1927.5 & 0 \cr
         17 & 1937.8 & 1937.5 & 4 \cr
         18 & 1947.5 & 1947.5 & 0 \cr
         19 & 1958.3 & 1957.5 & 10 \cr
         20 & 1968.3 & 1970.0 & -20 \cr
         21 & 1980.0 & 1980.3  & -4 \cr
         22 & 1989.6 & 1990.0 & -4 \cr
         23 & 2000.4 & 2000.8 & -5 \cr
         24 & 2013.0 & 2014.1 & -13 \cr
         25 & 2023.5 & 2024.8 & -17 \cr
         \hline
    \end{tabular}
    \label{T1}
\end{table}

\section{Does the Hemispheric Symmetry Indeed Exist?}

\begin{figure}    
\includegraphics[width=0.45\textwidth]{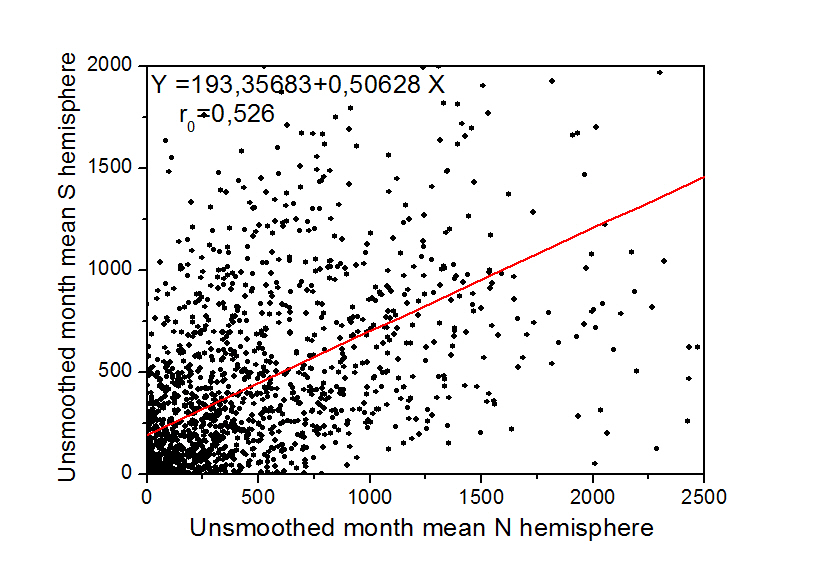}
\includegraphics[width=0.45\textwidth]{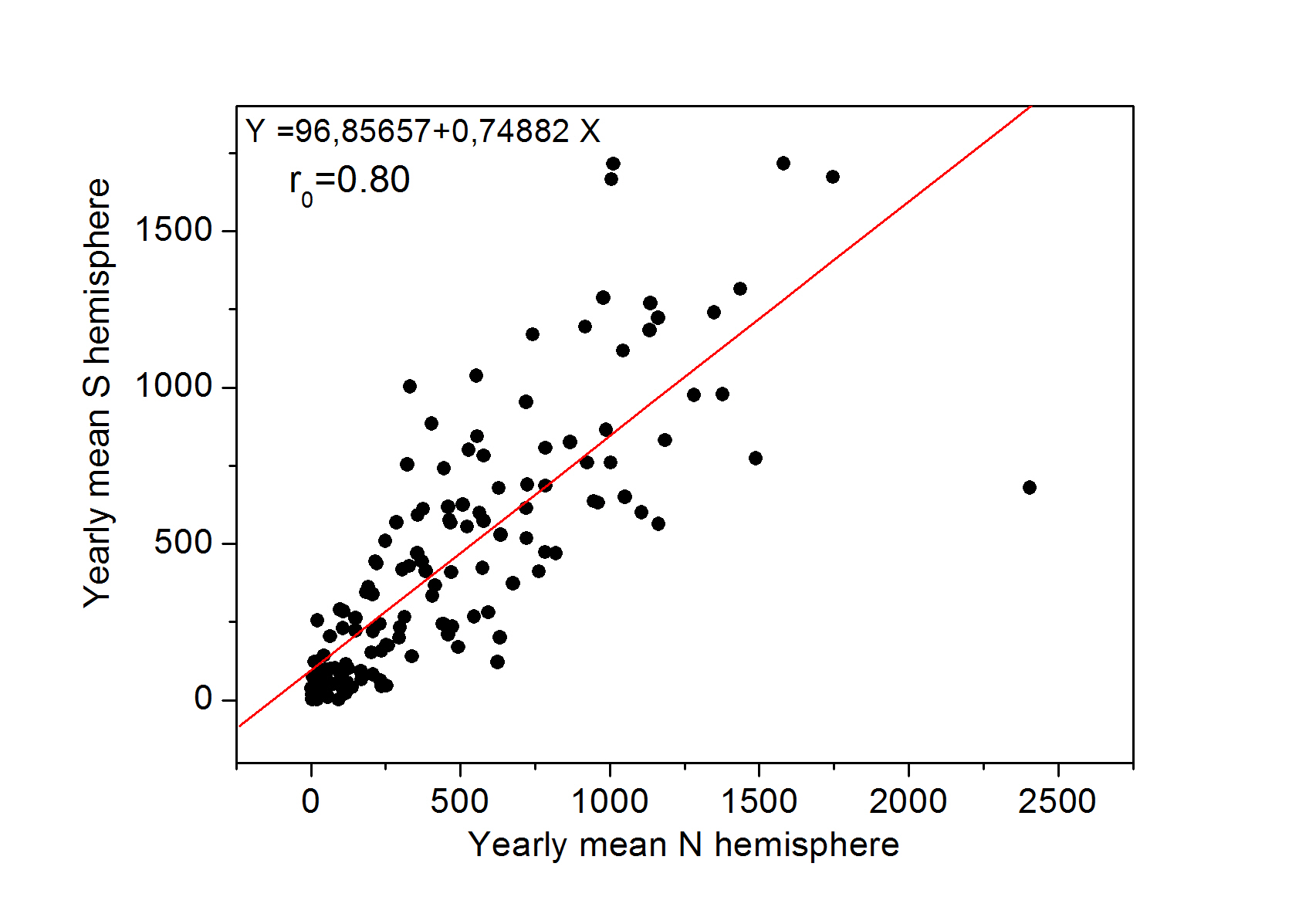}

\small
        \caption{Correlations between the monthly mean sunspot areas in the northern (vertical axis $Y$) and southern (horizontal axis $X$) hemispheres ({\it left panel} -- monthly data, {\it right panel} -- mean yearly data). Here $r_0$ is the correlation coefficient.
       }
\label{F4}
\end{figure}

Here, we are faced with an awkward question: Does the symmetry of the hemispheres indeed exist, or do we see only independent waves of activity in each hemisphere? To answer this question, we used the correlation of monthly data in the northern and southern hemispheres (Figure~\ref{F4}). In the left panel, the correlation is rather weak (for example, $r_0 = 0.526 \pm 0.0206$ with the number of pairs of points n= 1709). There is no apparent connection between $Y$ and $X$ at least in regions below 500 m.p.h. Recall that the modern concept of the solar cycle does not presume the physical link between the quantities connected by the solar dynamo mechanism. The correlation becomes substantially better for mean yearly data (e.g., $r_0=0.80 \pm 0.0051$  n=140), and the dots are much more concentrated around the line. The smoothing effect is mainly pronounced for the large $X$ and $Y$. In addition, we see that in the case of large sunspot areas the quantities in the northern hemisphere are substantially greater than in the southern.

\begin{figure}    
\includegraphics[width=0.5\textwidth]{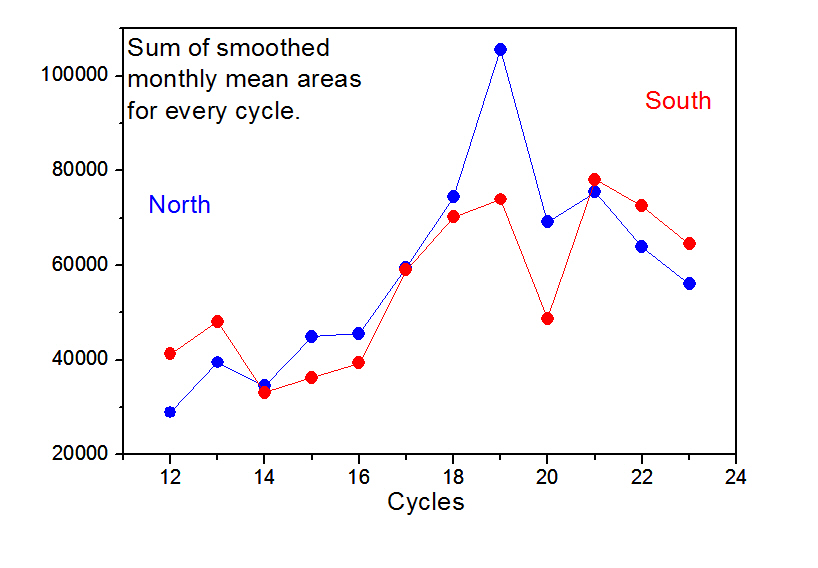}
\small
        \caption{The sum of smoothed monthly mean sunspot areas
        (in m.p.h.) for the North ({\it blue}) and South ({\it red}) hemisphere depending on the cycle number.}
\label{F5}
\end{figure}

Now, let us consider the total area of sunspots in a sunspot cycle (Figure~\ref{F5}). This quantity was used by Gnevyshev and Ohl to establish the well-known rule in 1948 (published in Russian only). For the current state of the problem, see \textcite{Netal24}. The quantity determines the total energy of the cycle and seems to be optimal to illustrate the spot production at the Sun.

One can see that there is a similarity between the North and South hemispheres in Figure~\ref{F5}. The correlation coefficient is not very small ($r_0 = 0.7978 \pm 0.167$, n=13). The South hemisphere is more active in Cycles 12, 13, while the North hemisphere is more active in Cycles 14 -- 20. In Cycles 21 -- 23,
the South hemisphere becomes more active again. According to the sunspot number data, it remains more active in Cycle 24 and in the current Cycle 25. Both hemispheres behave similarly, increasing or decreasing activity with the cycle number and follow the Gnevyshev-Ohl rule.

The north-south correlation coefficients are given separately for each cycle in Table~\ref{T2}. Again, we see that smoothing increases the correlation substantially.

\begin{table}
\caption{Correlation coefficient $r_0$ between North and South in sunspot areas.}
    \begin{tabular}{|c|c|c|}
    \hline 
Cycle number & $r_0$ for unsmoothed data & $r_0$ for  smoothed data \cr
       \hline
         12 & 0.47911 & 0.7831  \cr
         13 & 0.51107 & 0.8349  \cr
         14 & 0.41020 & 0.74291 \cr
         15 & 0.49346 & 0.89097  \cr
         16 & 0.48845 & 0.88160 \cr
         17 & 0.38249 & 0.78265  \cr
         18 & 0.46870 & 0.84788 \cr
         19 & 0.56158 & 0.81016 \cr
         20 & 0.38895 & 0.70511  \cr
         21 & 0.49183 & 0.85350  \cr
         22 & 0.56924 & 0.93810  \cr
         23 & 0.62778 & 0.89720  \cr
         \hline
    \end{tabular}
    \label{T2}
\end{table}

Summarizing data from Figures \ref{F4}, \ref{F5} and Table \ref{T2}, we see that the 11-year cycle indeed looks quite similar in both hemispheres except for some minor details. In order to separate these details we use the wavelet approach, i.e. a {\bf local version}
of the Fourier transform.
We calculate the total wavelet spectrum. First, let us calculate the wavelet coefficients.

\begin{equation}
W(a,t_0)=\frac{1}{|a|^{1/2}}\int\limits_{-\infty}^{\infty} 
f(t)\psi^{\ast} \left(\frac{t-t_0}{a}\right) {\mathrm d}t\, ,
\end{equation}
where 
 \begin{equation}
\psi (t)=e^{-t^2/\alpha^2}\left[e^{ik_0 t}-e^{-k_0^2 \alpha^2/4}\right],
\end{equation}
is the Morlet wavelet, i.e. Gaussian wave pocket, $\ast$ denotes complex conjugation, $a$ is the time scale and $W(a, t_0)$ is spectral density on the scale $a$ at the instant $t_0$ and $\alpha^2=2,\, k_0=2\pi$. 
The integral wavelet spectrum, i.e. an adaptation of the Fourier spectrum, is
 
\begin{equation}
S(a)=C_{\psi}^{-1/2}\int\limits_{-\infty}^{\infty} |W(a,t_0)|^2 {\mathrm d} t_0,
\end{equation}
where $C_{\psi}$ is 
\begin{equation}
C_{\psi}=\int\limits_{-\infty}^{\infty} |\omega|^{-1}|\hat{\psi}(\omega)|^2 {\mathrm d}\omega
\label{integral}
\end{equation}
and 
\begin{equation}
\hat{\psi}(\omega)=\int\limits_{-\infty}^{\infty} \psi(t)e^{-i\omega t}{\mathrm d} t
\end{equation}
is the Fourier transform of the wavelet. In fact, the integration in Equation~(\ref{integral}) is performed over time intervals when the analyzed signal is known. When comparing the spectra of the signals known in certain time intervals, it is necessary to normalize the spectra by the length of the analyzed interval.

\begin{figure}    
\includegraphics[width=0.7\textwidth]{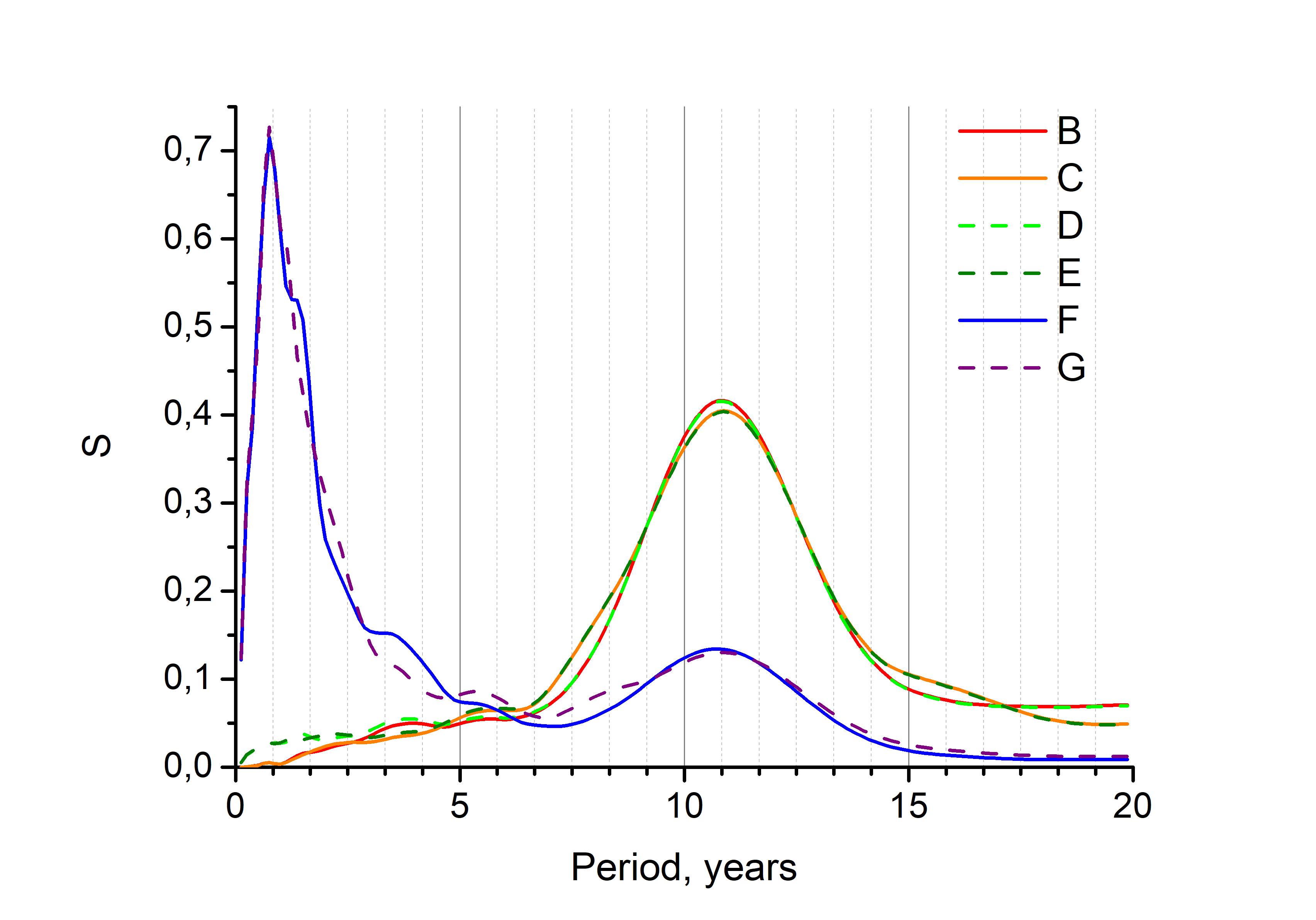}
\small
        \caption{Wavelet spectra for sunspot data, smoothed and unsmoothed data, northern and southern hemispheres.}
\label{F6}
\end{figure}

The integral wavelet spectra for sunspot data in the North and South hemispheres are shown in Figure~\ref{F6}.

We conclude from Figure~\ref{F6} that the 11-year cycles in both hemispheres are virtually identical in their spectral shape. The 11-year oscillations may be slightly shifted in time in one hemisphere with respect to the other. The other shape differences have timescales less than a year and can be considered as driven by processes different from the mean-field dynamo responsible for the main cycle.

Here, the following natural question arises: What about the mean surface magnetic field? We understand that the dynamo theory deals with the mean magnetic field somewhere inside the solar convection zone, while the observations provide information about the surface magnetic field.

To solve this problem, we used data of the surface mean-field observations with a resolution of 2 arcsec carried out at the John Wilcox Solar Observatory during almost 60 years. The rough data in the form of synoptic maps taken from \url{wso.stanford.edu/synopticl.html} were used. Original maps are accessible from Carrington rotation 1642 (starting on 27 May 1976) until rotation 2293 (starting on 6 January 2025). Each original map is a matrix with 30 lines equidistributed with respect to $\sin \phi$, where $\phi$ is the solar latitude). Each line contains 73 magnetic-field measurements distributed with a step of $5^\circ$ in longitude. 

\begin{figure}    
\includegraphics[width=0.7\textwidth]{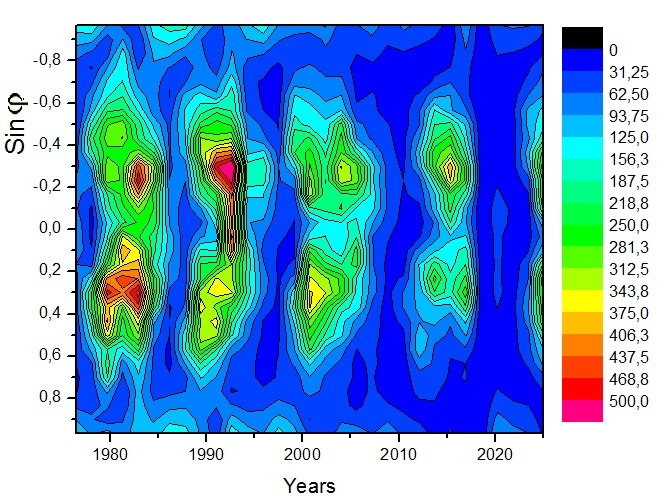}
\small
        \caption{Supersynoptic map of the mean absolute values calculated for each latitude and each rotation.}
\label{F7}
\end{figure}

Figure~\ref{F7} represents a supersynoptic map of the mean absolute field values calculated for each latitude and each rotation. As can be seen in the figure, after averaging, no pronounced asymmetry of the hemispheres is observed. Only the 11-year pattern is clearly visible. This confirms that the 11-year cycle is determined by long-lived large-scale structures associated with the mean magnetic field.

A time shift of the hemispheres may arise, but it does not exceed 10--20 months. Of course, the resolution of 30 arc minutes is still not low enough and individual active regions are visible even with such averaging. However, their time scale does not exceed several rotations.

\section{Small-scale Structures}

Now, let us turn to the small-scale structures in the data under discussion. We quantify them in terms of the following quantities.

\begin{equation}
\Delta N =S_{\mathrm N}- S_{\mathrm NSM}, \quad \quad \Delta S =S_{\mathrm S}- S_{\mathrm SSM},
\label{Del}
\end{equation}

i.e., the differences between the measured total sunspot areas in the North (N) and South (S) hemisphere and its smoothed values (marked NSM and SSM, respectively). We plotted the data for 1697 Carrington rotations. The difference $\Delta N$ changes sign 678 times, i.e. on average each 2.502 rotation. $\Delta S$ changes sign 715 times, i.e. on average each 2.364 rotations. Following \textcite{BO17}, we plotted histograms for the time intervals between sign changes to see if they were consistent with the interpretation of sign changes as a Poisson stream of events.
 
\begin{figure}    
\includegraphics[width=0.85\textwidth]{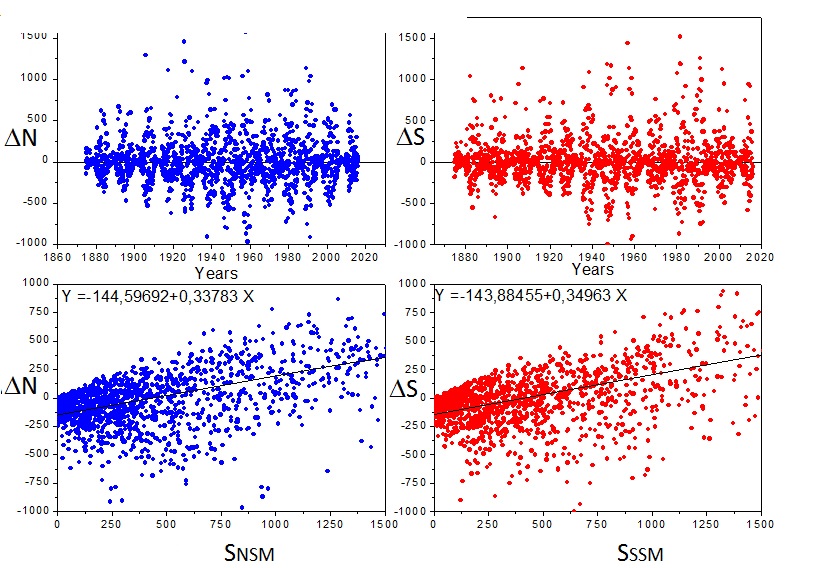}
\small
        \caption{Statistics of positive and negative deviations from the smoothed data on the timescale less than a year; {\it top} - hystogram, {\it bottom} - scattering field.}
\label{F8}
\end{figure}

Figure~\ref{F8} shows the time variation of $\Delta N$ (left column) and $\Delta S$ (right column) (upper row) and the smoothed sunspot area (lower row). One can see that the differences are more or less uniformly distributed in time and the distribution in the South hemisphere is more or less the same as in the North hemisphere (upper panels in Figure~\ref{F8}). The differences increase slightly with the amplitude of solar activity (the corresponding correlations are shown on the lower panels of Figure~\ref{F8}).

\begin{figure}    
\includegraphics[width=0.49\textwidth]{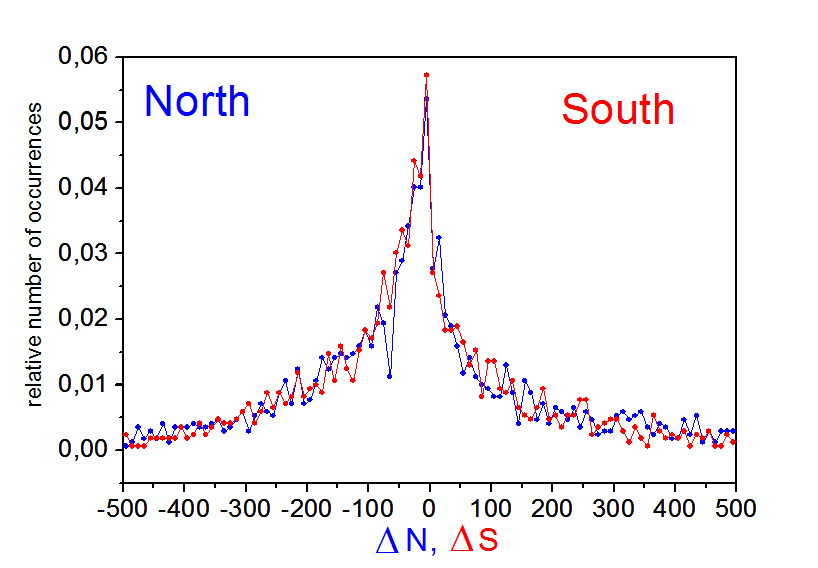}
\includegraphics[width=0.49\textwidth]{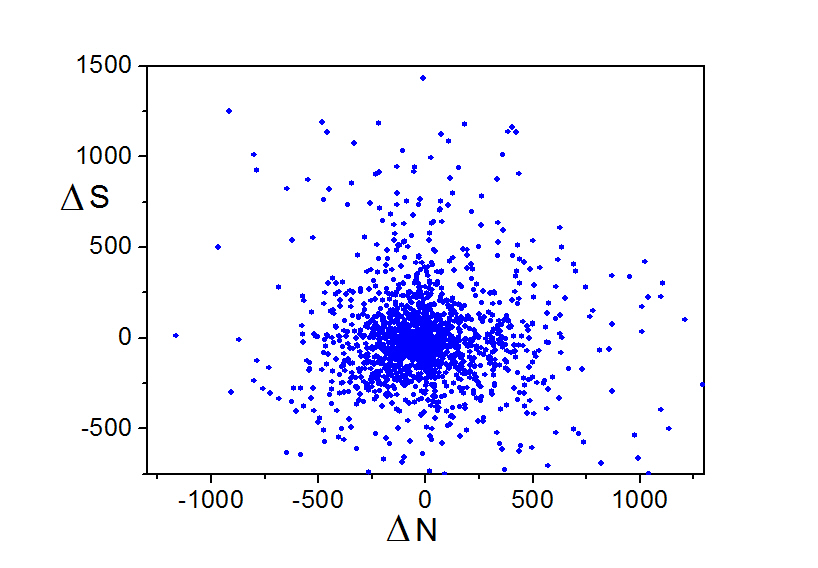}

\small
        \caption{Distribution of small-scale features: {\it left} - probability histogram (abscissa is $\Delta N$, blue and $\Delta S$, ordinata is the relative number of occurrences), {\it right} - scattering field.} 
\label{F9}
\end{figure}

The probability distribution of positive and negative deviations from the smoothed data on the time scale less than a year is shown in Figure~\ref{F9} (left). The amplitude of deviations does not exceed 500 m.p.h. We see from this figure that the positive and negative deviations from the smoothed data have a similar probability distribution (left panel); however, the positive and negative deviations look independent of each other. They give a symmetric spot on the scattering plane (right panel). We accept Figure~\ref{F9} as an indication that small-scale details in both hemispheres are produced by similar but independent processes.

We confirm the idea of independence of small features in the north and south hemispheres by two additional statistical tests. First, the signs of deviations in both hemispheres coincide in 51.9\% of cases (881 of 1697 cases) and differ in 48.1\% (816 of 1697 cases), which is consistent with the idea. The cases are evenly distributed over time. Second, we calculate the correlation coefficient between $\Delta N$ and $\Delta S$, which turns out to be quite small, only 0.0482.

\begin{figure}    
\includegraphics[width=0.49\textwidth]{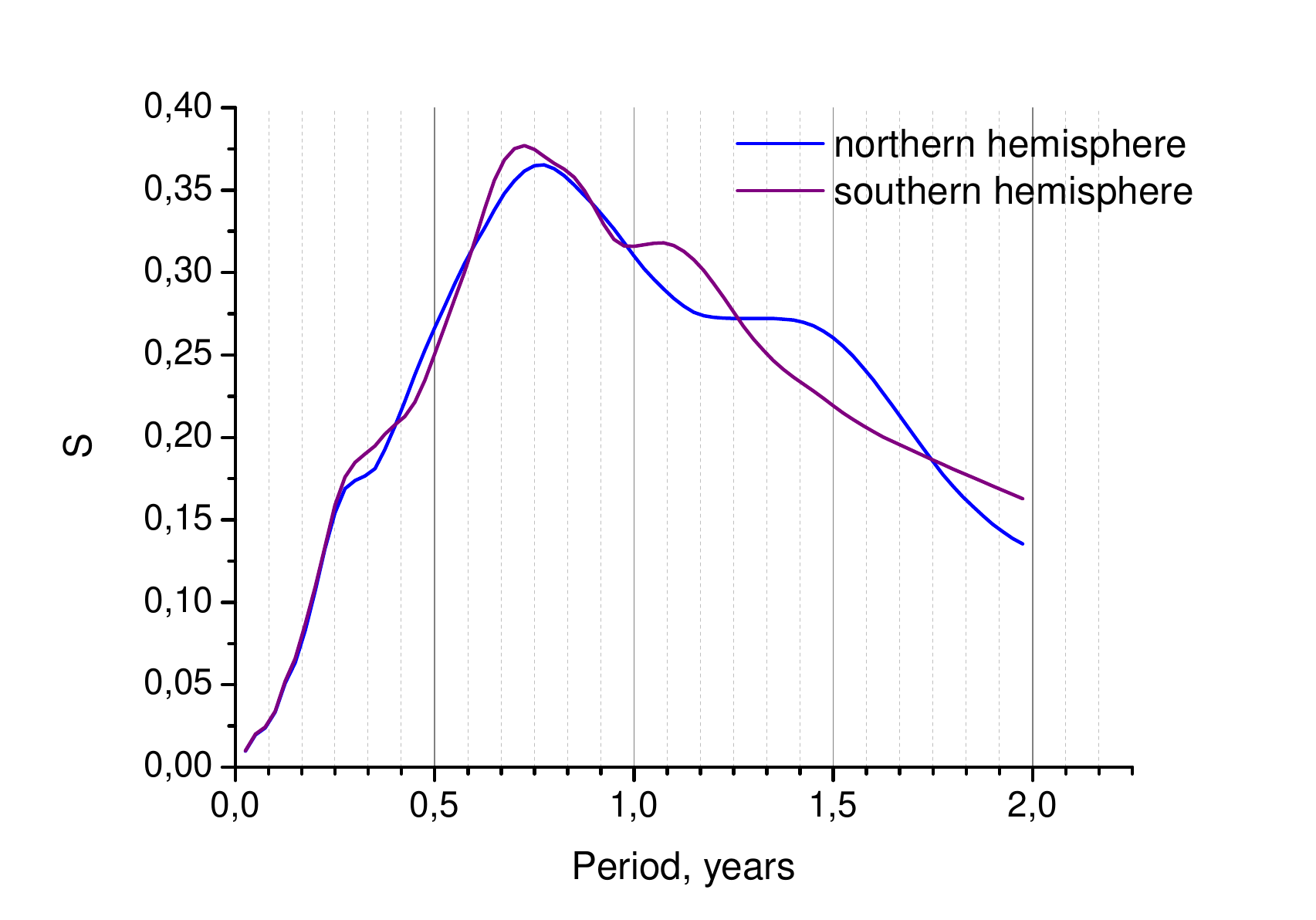}
\small
        \caption{Wavelet spectra for small-scale features, {\it blue} - North, {\it red} - South.}
\label{F10}
\end{figure}

And finally, we calculate the wavelet spectra for small-scale details for each hemisphere separately (Figure~\ref{F10}). The similarity of both spectra confirms our point.

\section{Conclusion and Discussion}

In summary, we show that surface magnetic field features such as sunspots and surface mean magnetic-field structures behave fundamentally differently in two time domains, i.e. on time scales shorter than a year and longer than a year. Short-time structures are distributed independently in the north and south hemispheres and present a random (Poisson-like) probability distribution over. In contrast, the long-time structures are (anti)symmetric with respect to the solar equator and follow the 11-year cycle. The symmetry is not fully perfect, but is well recognizable in all data under discussion. 

We can not directly observe the separation between the structures in the solar interior; however, it seems reasonable to expect that the separation is not a purely surface phenomenon, but reflects the corresponding separation at least in the upper part of the convective zone. We believe that these two types of magnetic structures are driven by two specific mechanisms, i.e. large-scale solar dynamo for the long-time structure and spot production for the short-time structures. 

In this paper, developing the idea of \textcite{BO17}, we analyze the statistics of time intervals between magnetic-field reversals in small-scale structures. We find that the reversals can be considered as a Poisson flow of events and arrive at the conclusion that the evolution of the magnetic field must include stochastic processes. Random fluctuations of dynamo drivers play an important role in some solar dynamo models and can explain cycle-to-cycle variations of solar activity in the framework of the Waldmainer rules (e.g. \cite{Petal12}) or even explain solar Grand minima, \textcite{Metal08}. 

Fluctuations in dynamo drivers can occur, in particular, in the Babkock-Leighton scheme. \textcite{GC09} 
demonstrated that such oscillations can lead to a magnetic configuration in which the poloidal magnetic field in one hemisphere is much larger than in the other.

\textcite{Detal06} used similar ideas in the context of the solar activity forecast. \textcite{Detal07} applied the idea of simulating the solar magnetic field to models with specific fluctuations in each hemisphere and found that the scheme gives realistic amplitude of activity cycles. The results are adequate if the smoothing is performed over at least 13 solar rotations and become inadequate when the smoothing time is shorter.   

Our conclusion ensures an observational support of the theoretical concept of separation between the mean magnetic field governed by mean-field dynamo equations and the small-scale processes that involve different physical mechanisms like the buoyancy and/or negative magnetic pressure or something else. 

Such separation was anticipated in different aspects in many papers. The novelty of our approach is that we demonstrate that isolation of these two cases is not a technical simplification but reflects the physical difference between the two processes. Of course, the processes are not fully independent, and the mean-field dynamo provides magnetic flux for sunspot activity. We see no need to insist that an additional dynamo activity in small scales independent on the 11-year cycle is impossible;  however, its contribution as available is difficult to separate from other processes (see, however, \textcite{Setal15}). On the other hand, small-scale structures sometimes propagate through the solar equator and produce some secondary link between the hemispheres \parencite{Oetal20}. As a whole, the results indicate an important role that processes in the leptocline play in the formation of the solar cycle. In particular, the results of \textcite{KR18, GK22} may be a message that the leptocline is just a layer responsible for the formation of short-term structures. 
 
The above results help to clarify the spatial structure of dynamo-driven stellar magnetic configurations using stellar butterfly diagrams \parencite[e.g.][]{BH07,Ketal10}. The point is that the problem requires fairly long-term (at least several years) monitoring of stellar activity to perform sufficient smoothing to obtain meaningful results.

The above results give an important message for the attempts to clarify spatial structure of dynamo driven stellar magnetic configurations using stellar butterfly diagrams \parencite[e.g.][]{BH07, Ketal10}. The message is that the problem requires quite a long stellar activity monitoring (at least several years) to perform sufficient smoothing to give meaningful results.

\section*{Acknowledgements}
All authors contributed to the study conception and design. 

Material preparation, data collection and analysis were performed by ASS under the guidance of VNO and DDS. 

The first draft of the manuscript was written by VNO and all authors commented on previous versions of the manuscript. 

All authors read and approved the final manuscript.

\section*{Data Availability}
All data on photospheric magnetic fields are available at
\url{http://wso.stanford.edu/synopticl.html}
and Sunspot data from the World Data Center SILSO, Royal Observatory of Belgium, Brussels. Version V2. \url{https://sidc.be/SILSO/datafiles}


\printbibliography

\end{document}